\documentclass[lettersize,journal]{IEEEtran}
\usepackage{amsmath,amsfonts}
\usepackage{algorithmic}
\usepackage{algorithm}
\usepackage{array}
\usepackage[caption=false,font=normalsize,labelfont=sf,textfont=sf]{subfig}
\usepackage[justification=centering]{caption}
\usepackage{textcomp}
\usepackage{stfloats}
\usepackage{url}
\usepackage{verbatim}
\usepackage{graphicx}
\usepackage{cite}
\usepackage{tabularx}
\usepackage{tabulary}
\usepackage{nicematrix}
\usepackage{array} % Required for vertical centering

\usepackage{placeins}

\usepackage{tikz}

\newcommand\copyrighttext{ \footnotesize \textcopyright{} 2025 IEEE.  Personal use of this material is permitted.  Permission from IEEE must be obtained for all other uses, in any current or future media, including reprinting/republishing this material for advertising or promotional purposes, creating new collective works, for resale or redistribution to servers or lists, or reuse of any copyrighted component of this work in other works.}
\newcommand\copyrightnotice{%
\begin{tikzpicture}[remember picture,overlay]
\node[anchor=south,yshift=6pt] at (current page.south) {\fbox{\parbox{\dimexpr\textwidth-\fboxsep-\fboxrule\relax}{\copyrighttext}}};
\end{tikzpicture}%
}

\begin{document}

\title{A Primer on AP Power Save in Wi-Fi 8: Overview, Analysis, and Open Challenges}

\author{\IEEEauthorblockN{Roger Sanchez-Vital\thanks{Roger Sanchez-Vital (corresponding author), Carles Gomez, and Eduard Garcia-Villegas are with Universitat Politecnica de Catalunya;}%\IEEEauthorrefmark{1}
, Andrey Belogaev\thanks{Andrey Belogaev and Jeroen Famaey are with University of Antwerp and imec}%\IEEEauthorrefmark{2}
, Carles Gomez%\IEEEauthorrefmark{1}
, Jeroen Famaey%
%\IEEEauthorrefmark{2}
, and Eduard Garcia-Villegas%\IEEEauthorrefmark{1} %Decide the author order
        % <-this % stops a space
%\thanks{This paper was produced by the IEEE Publication Technology Group. They are in Piscataway, NJ.}% <-this % stops a space
%\thanks{Manuscript received April 19, 2021; revised August 16, 2021.}
}
%\IEEEauthorblockA{\IEEEauthorrefmark{1}Department of Network Engineering,  Universitat Politecnica de Catalunya\\
%Email: \{roger.sanchez.vital, carles.gomez, eduardo.garcia\}@upc.edu\\}
%\IEEEauthorblockA{\IEEEauthorrefmark{2} IDLab, University of Antwerp -- imec, Belgium \\ 
%Email: \{andrei.belogaev, jeroen.famaey\}@uantwerpen.be}
}



\maketitle

\copyrightnotice

\begin{abstract}
Wi-Fi facilitates the Internet connectivity of billions of devices worldwide, making it an indispensable technology for modern life. Wi-Fi networks are becoming significantly denser, making energy consumption and its effects on operational costs and environmental sustainability crucial considerations. Wi-Fi has already introduced several mechanisms to enhance the energy efficiency of non-Access Point (non-AP) stations (STAs). However, the reduction of energy consumption of APs has never been a priority. Always-on APs operating at their highest capabilities consume significant power, which affects the energy costs of the infrastructure owner, aggravates the environmental impact, and decreases the lifetime of battery-powered APs. IEEE~802.11bn, which will be the basis of Wi-Fi 8, makes a big leap forward by introducing the AP Power Save (PS) framework. In this article, we describe and analyze the main proposals discussed in the IEEE~802.11bn Task Group (TGbn), such as Scheduled Power Save, (Semi-)Dynamic Power Save, and Cross-Link Power Save. We also consider other proposals that are being discussed in TGbn, namely the integration of Wake-up Radios (WuRs) and STA offloading. We then showcase the potential benefits of AP PS in several scenarios, including a deployment of 470 real APs in a university campus. Our numerical analysis reveals that AP power consumption can be decreased on average by up to 28 percent, with further improvement potential. Finally, we outline the open challenges that need to be addressed to optimally integrate AP PS in Wi-Fi and ensure its compatibility with legacy devices.
\end{abstract}


\section{Introduction}
\IEEEPARstart{W}{i-Fi} is everywhere, providing high-speed wireless connectivity to homes, office buildings, university campuses, airports, and industrial sites. The IEEE~802.11bn Task Group (TGbn) aims to tackle the access point (AP) power consumption issue in the upcoming Wi-Fi 8, with a new Power Save (PS) feature~\cite{primer_wifi8}. Its objective is threefold: 

\begin{itemize}
    \item To enhance the lifetime of battery-limited devices acting as APs, for example, mobile soft APs.
    \item To cut down on the energy bills of Wi-Fi network infrastructure.
    \item To reduce the environmental impact of dense Wi-Fi network deployments.
\end{itemize}

The energy consumption of Wi-Fi network infrastructure is a growing concern, as the equipment is constantly turned on and works mostly at its highest capabilities to provide the best Quality of Service (QoS). This is not energy-efficient, as the load of an AP typically follows predictable temporal usage patterns that present clear opportunities for energy savings~\cite{ap_dataset}. In contrast to cellular networks~\cite{green_cellular_survey}, AP power saving has never been a priority for Wi-Fi. However, now it is seen as a crucial issue. WIK-Consult estimated that, if there was at least one Wi-Fi AP at every EU household, the yearly power consumption of such devices could reach 26,640 GWh, leading to 6,069 million metric tons of carbon emissions~\cite{wifi_ap_consumption}. Dense office and campus Wi-Fi deployments further exacerbate the issue.
While AP energy consumption has not been a priority, past IEEE~802.11 standards have put a lot of emphasis on station (STA) power saving. Even the first Wi-Fi standards already allowed STAs to sleep if no downlink (DL) frames were buffered. More recently, IEEE 802.11ah and later 802.11ax proposed Target Wake Time (TWT), allowing STAs to perform scheduled sleep cycles in between uplink (UL) and DL transmissions \cite{twt_ec}. Table~\ref{tab:power_saving_mechanisms} provides a comparison of the main STA-side PS mechanisms defined in the standard.

\begin{table*}[htb]
    \centering
    \caption{Comparison of principal IEEE 802.11 STA-side Power-Saving Mechanisms \cite{energy_saving_wifi}}
    \resizebox{\textwidth}{!}{%
    \begin{tabularx}{\linewidth}{|>{\centering\arraybackslash}m{3cm}|X|>{\centering\arraybackslash}m{1.5cm}|X|>{\centering\arraybackslash}m{1.5cm}|}
        \hline
        \textbf{STA PS Mechanism} & \textbf{Functionality} & \textbf{Traffic direction} & \textbf{Power-Saving Effect} & \textbf{Amendment} \\
        \hline
        \hline
        Traffic Indication Map (TIM) & Indicates buffered unicast data in Beacon frames & DL & STAs sleep for multiple Beacon intervals, and stay awake only when there is buffered data & Legacy \\
        \hline
        Power Save Poll (PS-Poll) & Enables STAs to request for buffered frames (one per each frame) & DL & STAs poll for buffered data upon TIM indication and sleep in between & Legacy \\
        \hline
        Automatic Power Save Delivery (APSD) & Enables STAs to request for buffered frames (multiple at once, at scheduled appointments) & DL & STAs poll for buffered data upon TIM indication and sleep in between & 802.11e \\
        \hline
        Power Save Multi-Poll (PSMP) & Coordinates delivery times for uplink and downlink traffic & DL \& UL & Efficient scheduling of packet transmissions & 802.11n (obsolete) \\
        \hline
        Spatial Multiplexing Power Save (SMPS) & Dynamic utilization of multiple antennas & DL \& UL & STAs can utilize fewer antennas to save power & 802.11n \\
        \hline
        Wireless Network Management (WNM) Sleep Mode & STAs can negotiate extended sleep periods & DL & Increases sleep time & 802.11v \\
        \hline
        Transmission Opportunity (TXOP) Power Save & Allows STAs to go to sleep until the TXOP is completed to save power & DL \& UL & Reducing idle time inside TXOP & 802.11ac\\
        \hline
        Restricted Access Window (RAW) & Restricts access to the medium to specific groups of STAs & UL & Reducing idle time & 802.11ah \\
        \hline
        Target Wake Time (TWT) & Schedules specific wake times for devices & DL \& UL & STAs can wake up only at scheduled time slots & 802.11ax (prev. 802.11ah)\\
        \hline
    \end{tabularx}}
    \label{tab:power_saving_mechanisms}
\end{table*}

In contrast to the plethora of STA-side power saving mechanisms, only a few simple options are available on the AP side. Specifically, the AP can reduce the Bandwidth (BW), the number of Spatial Streams (SS), or disable links in case it supports Multi-Link Operation (MLO). Such changes affect all the STAs associated with the AP, and cannot be dynamically negotiated. Additionally, the AP is not normally allowed to sleep, as it should be able to reply to Probe Requests of new STAs, and regularly send Beacons. Wi-Fi Direct defines PS mechanisms for one of the STAs, called Group Owner (GO), which assumes a role analogous to an AP~\cite{wifi_direct}. These mechanisms allow the GO to schedule sleep periods akin to Scheduled PS. However, to enhance AP power savings, TGbn is developing other novel mechanisms beyond those proposed for Wi-Fi Direct. Out of six mechanisms discussed in this article, only Scheduled PS and Wake-up Radios (WuRs) partially incorporate or adapt older Wi-Fi power-saving mechanisms, while the remaining four are new to Wi-Fi.

This article provides an overview and analysis of AP PS mechanisms. First, we overview the main proposals that have already passed motions in TGbn and therefore most probably will be featured in the upcoming Wi-Fi 8 standard. Second, we describe other notable proposals discussed in TGbn. Then, we present a case study evaluating the performance of one of the proposed mechanisms in terms of power consumption and energy savings. Finally, we highlight several crucial open challenges.

\section{AP Power Save in IEEE 802.11bn}

TGbn currently distinguishes four AP PS mechanisms, namely Scheduled, Dynamic, Semi-Dynamic, and Cross-Link. Below, we describe each of them in detail.

\subsection{Scheduled Power Save}
\label{subsec:scheduled_ps}

Scheduled PS intends to reduce the energy consumption by switching between the AP's states according to a schedule~\cite{considerations_ap_power_save}. Specifically, the AP can switch between five states with different power consumption profiles.

\begin{itemize}
    \item \textbf{Doze state:} The AP disables its radio interface. It is not able to transmit, receive, or listen.
    \item \textbf{Listen state:} The AP only performs Clear Channel Assessment (CCA). It can switch to receive data, but cannot transmit. 
    \item \textbf{Interruptible listen state:} The AP performs CCA, can receive data, and can quickly switch to transmission, e.g., to respond to legacy devices.
    \item \textbf{Reduced capabilities:} The AP reduces its capabilities to a more basic configuration to save energy, e.g., 20 MHz BW and 1 SS. It can both transmit and receive data.
    \item \textbf{Full capabilities:} The AP operates using its highest-performance parameters to provide the best QoS to users. It can both transmit and receive data.
\end{itemize}

The AP elaborates its own state schedule and disseminates it via Beacons, Probe Responses, or Action Frames. The schedule includes groups of periodic intervals, PS periods, and Service Periods (SPs), each mapped to a state. The term SP, inherited from TWT, are intervals where the AP prioritizes high-rate data exchanges. Signaling frames convey schedule details, such as interval duration and periodicity, power state information, and capabilities information (e.g., BW and SS count). The schedule information can be reused from TWT-related frames. STAs adjust states accordingly, like entering doze state during AP doze intervals. The STAs that missed the schedule due to being in doze state can request it by sending a Schedule Request. 

STAs can request the AP to change its schedule by sending a Presence Request. For instance, an STA that is aware of the AP's power-saving schedule can send a request to the AP to temporarily exit doze mode at a predetermined time to meet specific QoS requirements. At any point in time, the AP can autonomously change the schedule based on network conditions, traffic volume, desired power consumption reduction, QoS requirements, etc. If the schedule is changed for any reason, the AP must disseminate the new schedule and is not allowed to switch to a new schedule before all the STAs have had the opportunity to receive it, either via Beacons or in reply to Schedule Requests. 

To avoid potential issues, several factors must be considered. First, if any legacy STAs are associated, the AP should avoid going into doze state. Such precaution will prevent legacy STAs from losing association and severe packet losses. Second, the newly-associated and just-woken STAs should wait for Beacons or send Schedule Requests to obtain the latest AP schedule. Third, whenever the AP goes into doze state, backoff counters at both the AP and STAs should freeze. If they keep counting down, backoff may reach zero for multiple devices simultaneously upon AP wake-up, leading to collisions. However, neighboring networks can keep counting down, which may lead to fairness issues. Finally, sleep patterns should align with STAs' QoS requirements. 

\subsection{Dynamic Power Save}

Even without switching to the doze state, the AP can save energy by reducing its capabilities. More specifically, its energy consumption depends on BW, Number of Spatial Streams (NSS), Modulation and Coding Scheme (MCS), and number of active links (in case of MLO) being used during transmission, reception and listening. The current standard already allows the AP to change these parameters and announce the changes to the associated STAs. However, there are no mechanisms to dynamically change the AP's capabilities on demand, e.g., on request from an STA. Such a mechanism, called Dynamic Power Save (DPS), is introduced by TGbn~\cite{dps_operation}.

DPS generally keeps the AP in Low Capability Mode (LCM), a power-saving mode that restricts the AP to a basic configuration with reduced BW, NSS and MCS. In this state, the AP is able to: (i) transmit Beacons and management frames; (ii) listen to the channel; and (iii) communicate with legacy STAs, as well as with DPS-capable STAs whose data flows can be served without increasing the AP's capabilities. Once an STA intends to perform frame exchange at higher capabilities, e.g., for higher throughput or lower latency, it sends a special control frame to trigger the AP to change its capabilities, i.e., to switch to High Capability Mode (HCM).

For signaling, DPS reuses the same approach as Enhanced Multi-Link Single-Radio (EMLSR) operation, which appeared in Wi-Fi 7 as a type of MLO. In EMLSR, an Initial Control Frame (ICF) is sent by a multi-link AP to solicit an STA to switch to its secondary link with higher capabilities. To compensate for the switching delay and allow the STA to reply with an Initial Control Reply (ICR), padding is added to the ICF. ICR will be introduced in IEEE 802.11bn as a frame that signals the availability information of an STA. The concept is already present in Wi-Fi 7, but without a specific frame definition. In contrast to EMLSR, DPS is triggered by an STA, which means that an STA can request an AP to switch capabilities by sending an ICF, and ICR is sent by an AP in reply. To facilitate the AP to error check ICF even before reception of the padding, an intermediate FCS before the padding field has been proposed.

The unified frame format for ICF and ICR is still under discussion. For DPS, the ICF may indicate: (i) configuration (BW, NSS, MCS); (ii) HCM duration, either explicitly or via timeout (e.g., inactive period). LCM parameters can be indicated in the Capabilities Information field, recognizable by legacy devices. Besides, transition delays to switch capabilities are manufacturer-specific and should be disseminated to STAs via an IEEE 802.11bn specific Capabilities Information field value, so that the STAs can add a suitable amount of padding.

\subsection{Semi-Dynamic Power Save}

When multiple DPS-capable STAs are associated with an AP, each of them can solicit an AP to switch to HCM by sending an ICF. If an AP responds to every ICF, it cannot sustain LCM for an extended duration, which restricts the potential power-saving benefits of DPS. To address this issue, TGbn proposed Semi-Dynamic Power Save (SDPS) mechanism, which is a modified DPS mechanism where an AP can selectively react to ICFs based on traffic demands and power-saving requirements. When the AP decides not to react to an ICF immediately, it can still remember the request from the STA, and trigger the pending transmission later when it switches to HCM, e.g., by transmitting a Trigger Frame (TF). To force the switching to HCM for critical traffic, STAs can mark their ICFs with a special flag, e.g., requiring Low Latency (LL) traffic. SDPS can be combined with Scheduled PS to increase the performance. 

\subsection{Cross-Link Power Save}\label{sec:cross-link}

The Scheduled PS mechanism may negatively affect the performance of sporadic critical traffic, e.g., when it arrives during doze intervals. To mitigate such effects, TGbn proposes to exchange management information for other links via the active link between a pair of Multi-Link Devices (MLDs), depending on the needs of STAs~\cite{cross_link_operation}. In this case, an MLD AP maintains an active link that supports several functionalities, like discovery, active probing, and association for all types of STAs, including legacy STAs. Upon reception of a cross-link wake-up frame on its active link from an STA, the MLD AP can enable its other links. This way, the AP can offload this STA to another link with higher capabilities, or start serving it with multiple links.

The way this feature will be implemented is still under discussion. One possibility is to reuse the AP Assistance Request (AAR) control information field introduced in IEEE 802.11be. This field contains a bitmap of link identifiers that can be used as an indication of the links that are requested to wake up. However, currently it is used only in the DL to assist the MLD STA to recover its medium synchronization. Hence, wake-up request implementation will require additional changes in the standard.

\begin{figure*} 
  \centering
  \captionsetup[subfloat]{labelfont=footnotesize,textfont=footnotesize}
  \subfloat[\label{fig:throughput_mode}]{%
       \includegraphics[width=\linewidth]{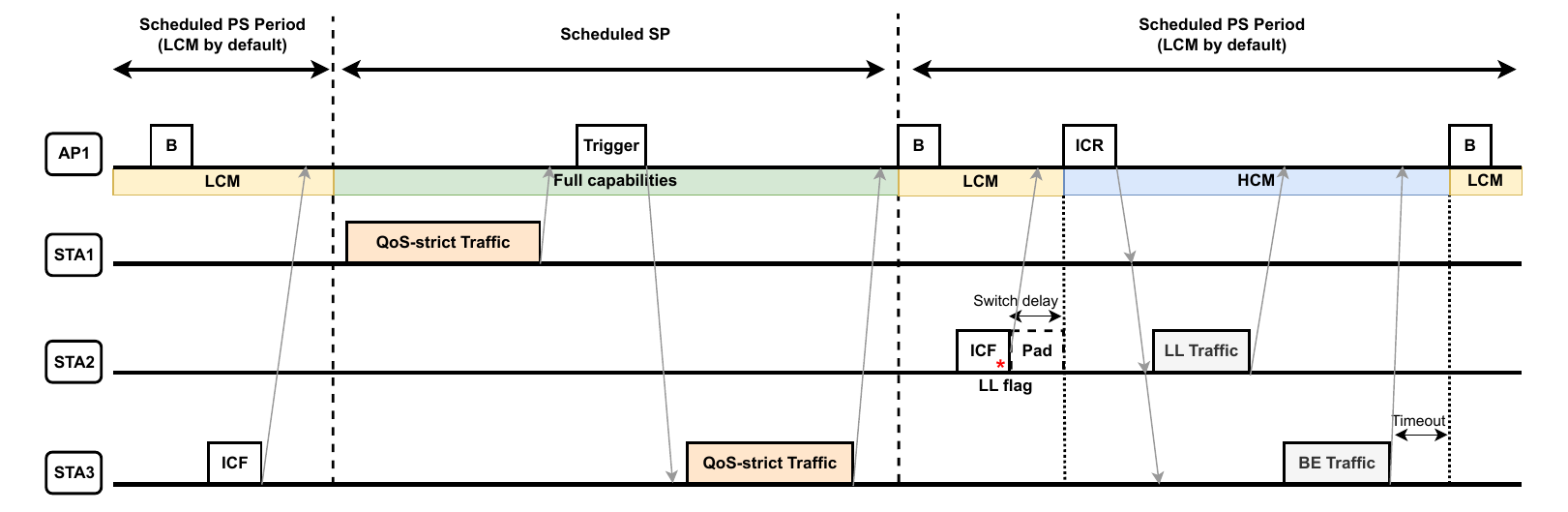}}
    \vfill
  \subfloat[\label{fig:energy_efficiency_mode}]{%
        \includegraphics[width=\linewidth]{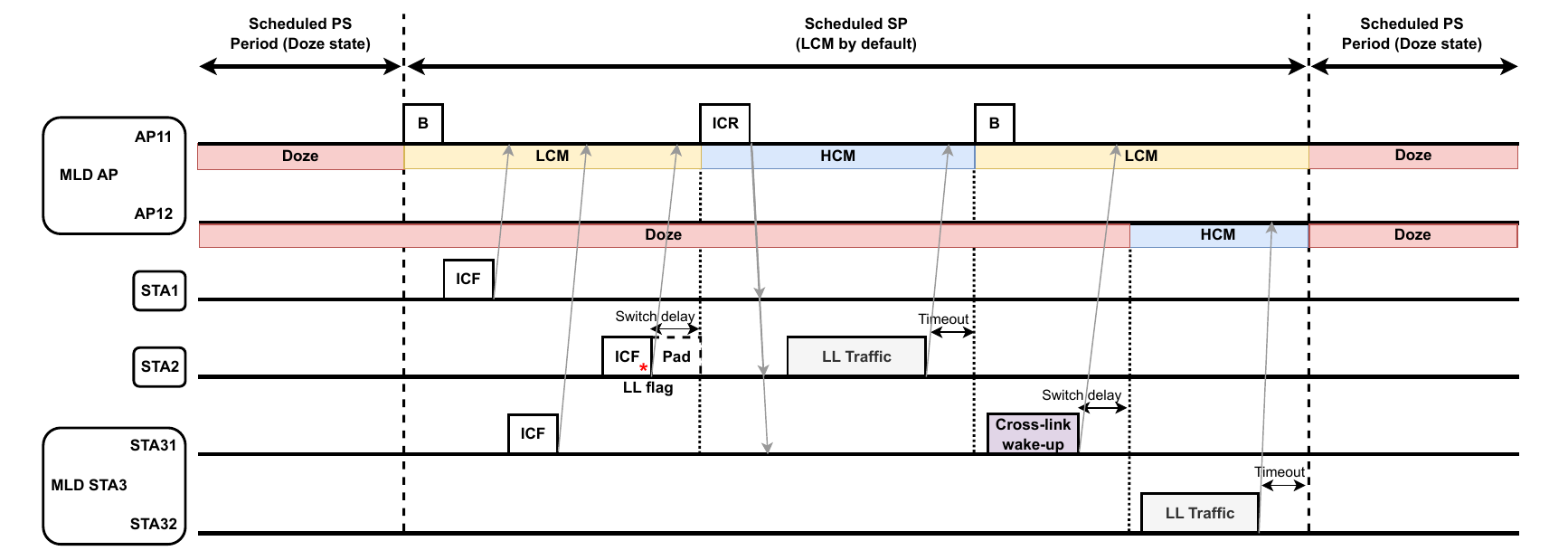}}
  \caption{Combinations of PS mechanisms: a) Type 1 (uninterrupted service); b) Type 2 (energy efficiency).}
  \label{fig:diagrams} 
\end{figure*}

TGbn also defines the possibility to combine PS mechanisms, specifically into two types: Type 1, in which the objective is to provide uninterrupted service. Therefore, in the SPs, the AP uses its full capabilities to provide maximal QoS, and transitions to SDPS in PS periods; and Type 2, in which the aim is to maximize the energy efficiency by operating in SDPS in the SPs while being in Doze state otherwise. Figs.~\ref{fig:throughput_mode} and~\ref{fig:energy_efficiency_mode} depict frame exchange sequences for Types 1 and 2, respectively. In Type 1, when the AP intends to save power by using DPS mode, the AP stays in LCM for energy efficiency, but switches to HCM for LL traffic requiring higher data rates. To request the capability switch, an STA sends an ICF frame which, after a padding time to let the AP interface switch the capabilities, is acknowledged by an ICR. Note that the AP can decide whether to switch its capabilities or not upon reception of an ICF. On the other hand, in Type 2 SPs, the AP uses DPS and Cross-Link PS. Best Effort (BE) traffic can be supported with both capability modes, but for time-critical traffic, MLDs can enable an additional link via a cross-link wake-up. Finally, during power-saving periods, the AP goes to sleep to maximize energy efficiency. Scheduled periods are disseminated via Beacons (B).

\begin{figure} 
  \centering
  \captionsetup[subfloat]{labelfont=footnotesize,textfont=footnotesize}
  \subfloat[\label{fig:pc_throughput}]{%
       \includegraphics[width=\linewidth]{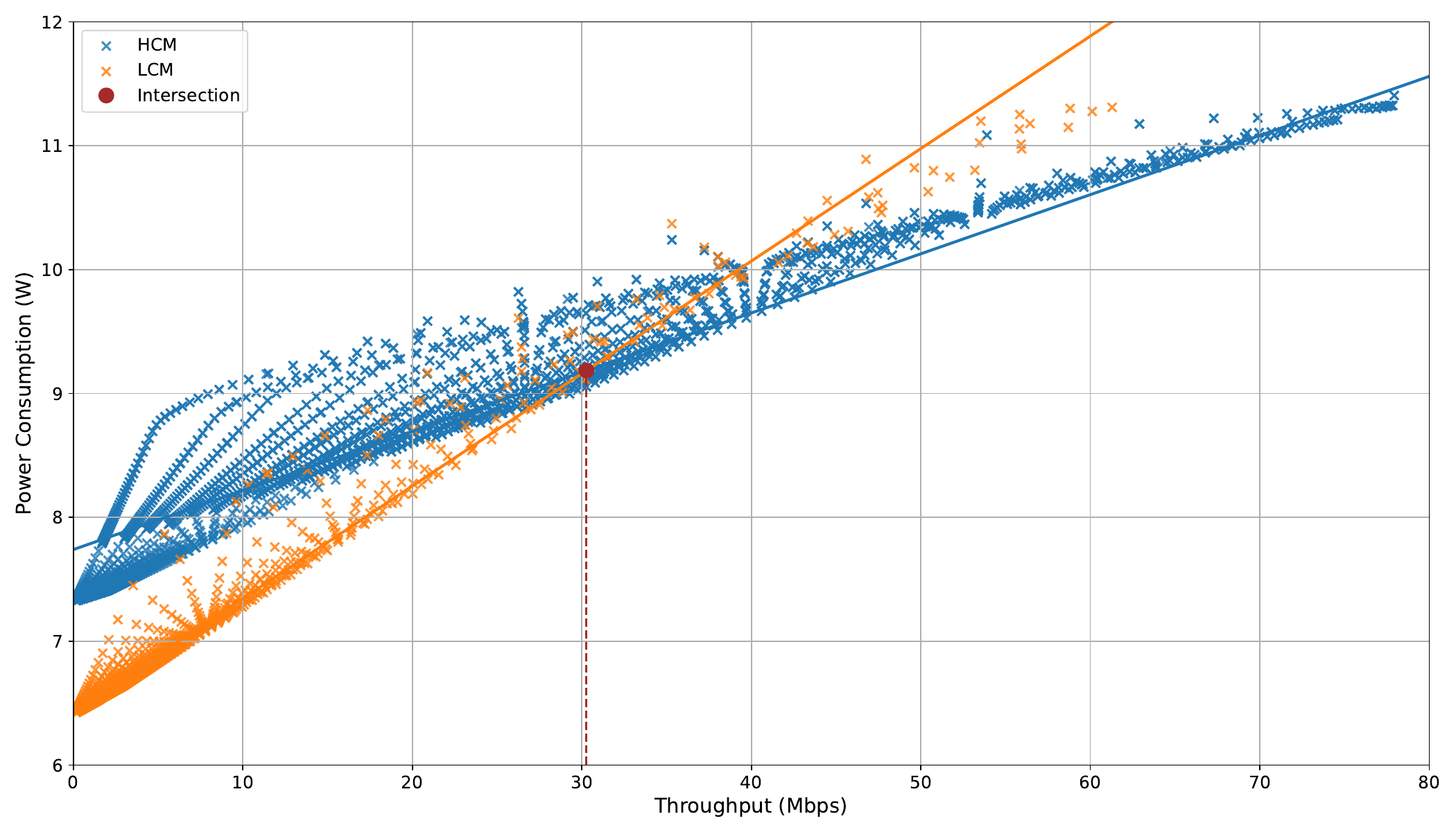}}
    \vfill
  \subfloat[\label{fig:delay_cm}]{%
        \includegraphics[width=\linewidth]{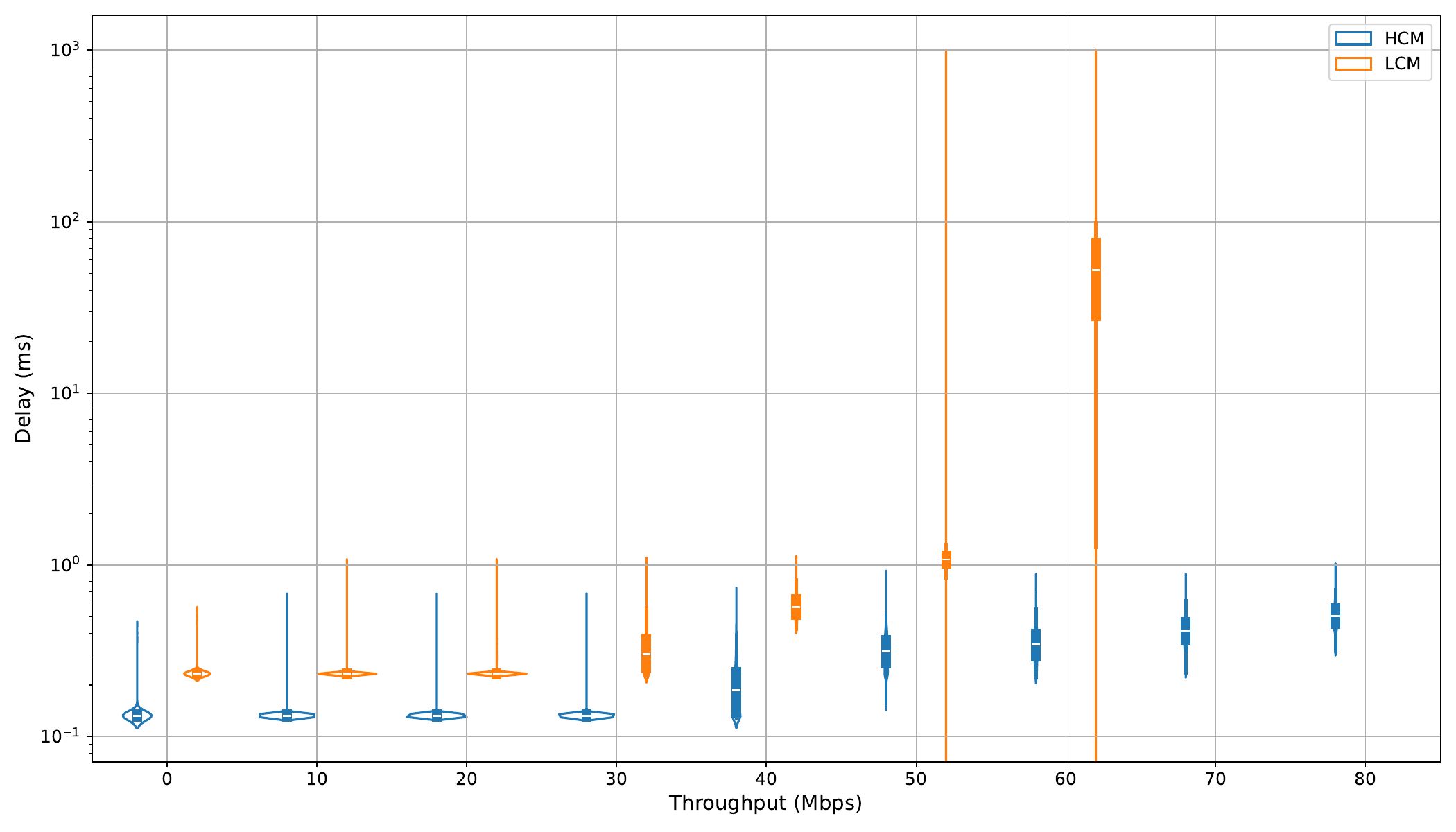}}
  \caption{Performance for LCM and HCM versus throughput: a) Power consumption; b) Per-packet delay.}
  \label{fig:dps} 
\end{figure}

\section{Other proposals}

Apart from the aforementioned features that have already been thoroughly discussed in TGbn and are on track to be added to the standard, some other proposals for AP-side power saving deserve attention. Those include the use of WuRs and the offloading of STAs to other APs.

\subsection{Wake-up Radios}

WuRs have arisen as a feasible way to maximize the energy efficiency of a device, while supporting spontaneous traffic. The IEEE 802.11ba amendment standardizes WuR operation in Wi-Fi networks. The main idea is to add a secondary low-power radio that stays active, allowing the power-hungry Primary Connectivity Radio (PCR) to go into doze state when idle. Unlike MLO, WuRs lack transmission capabilities and consume extremely low power, under $1$ mW, which allows an AP to keep it always on at minimal cost, while only waking up the PCR for application data exchange. To maximize efficiency, WuRs demodulate a simple On-Off Keying (OOK) waveform from another STAs' PCR, offering relatively low data rates. Therefore, it is primarily considered for control data, such as wake-up frames. 

A proposal has been made to integrate the WuR concept into 802.11bn~\cite{wur_11bn}, though it presents several challenges that must be addressed. First, serving urgent traffic becomes challenging, since transmitters either have to wait before PCR becomes available, or use the low-data-rate WuR. Although MLO can suffer from similar problems, an active MLD link can still be used for application data transmission. Second, WuRs use a different waveform than standard Wi-Fi, requiring thorough coexistence studies with legacy devices. Besides, a WuR may have a different communication range than its PCR. 

\subsection{STA offloading}

APs must ensure fair connectivity for all STAs. However, in dense deployments, the load is non-uniformly distributed between the APs. An underutilized AP (e.g., serving a single STA with sparse traffic), still consumes a lot of power. Literature suggests offloading STAs from such APs, allowing the AP to temporarily transition to a deep doze state~\cite{dynamic_ap_switching}.

Before Wi-Fi 8, the standard did not offer any framework for coordination between APs. Hence, an AP could only decide to stop serving an STA, but could not control the rest of the handover procedure. This caused data flow disruptions, packet loss and unnecessary retransmissions. For instance, an STA might select the wrong AP due to incomplete RSSI measurements. With the introduction of Multi-AP Coordination to TGbn discussion, Wi-Fi APs can now exchange information to facilitate seamless STA roaming.

\begin{figure*}[tbp]
\centerline{\includegraphics[width=\linewidth,trim = 3cm 0cm 0cm 0cm, clip]{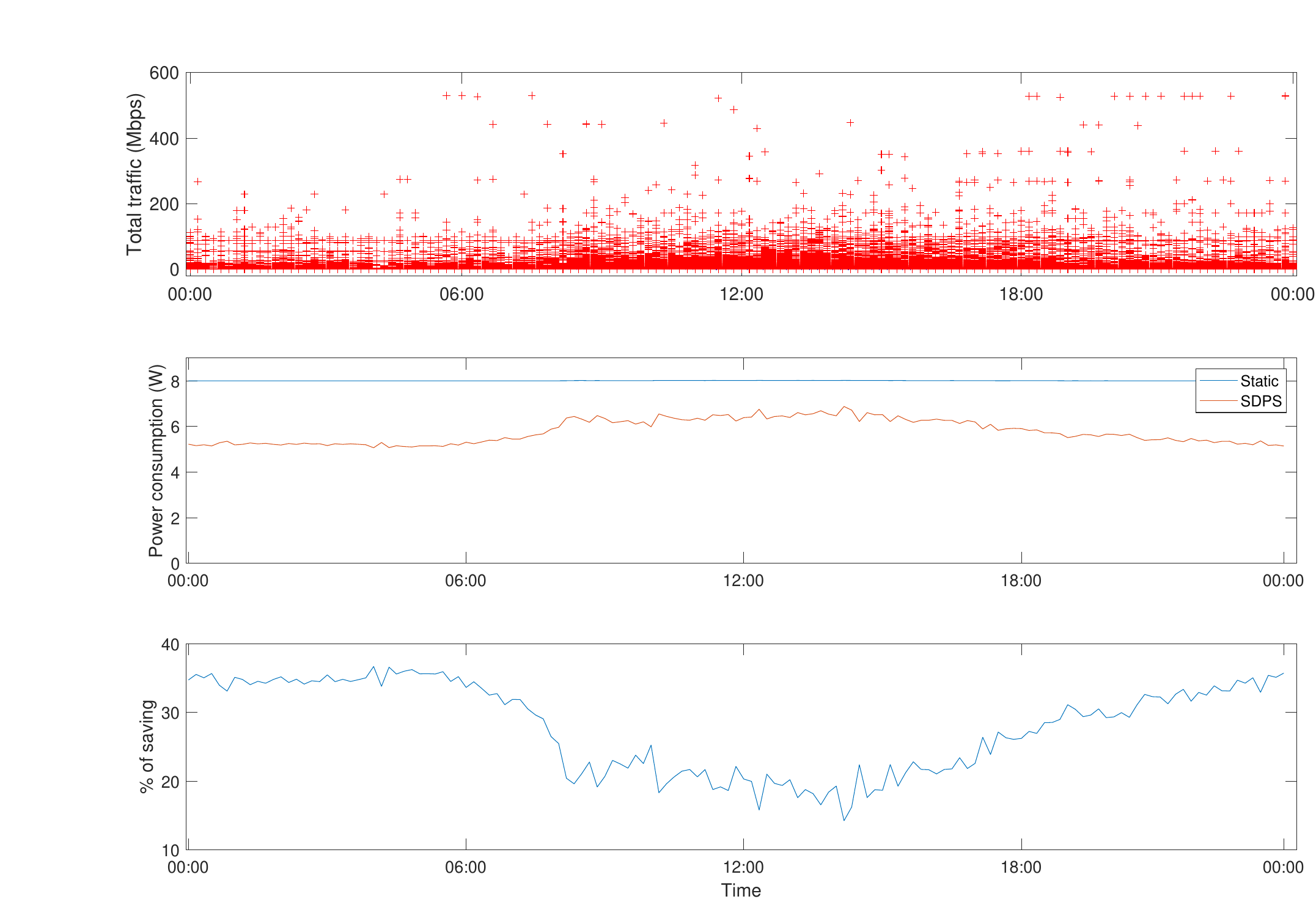}}
\caption{Total traffic, power consumption and SDPS power savings for 470 APs on a weekday based on dataset~\cite{ap_dataset}.}
\label{fig:es_dataset}
\end{figure*}

\section{Case study: Energy Saving Analysis}

This section outlines the potential energy savings of Wi-Fi~8 AP PS mechanisms. Scheduled PS is expected to have similar energy-saving effects as TWT, but targeting the AP instead of STAs. The impact of TWT is well-documented, with Yang \textit{et al.}~\cite{twt_ec} recently revealing it can reduce STA energy consumption by a factor of three. Since the AP also transmits Beacons and management frames, energy savings for AP-side Scheduled PS in a single-STA scenario will be slightly lower and decrease with more STAs. Further investigation of Scheduled and Cross-Link PS is outside the scope of this study.

Let us consider the DPS approach. Fig.~\ref{fig:dps} provides a comparison between HCM and LCM in terms of power consumption and per-packet delay as a function of application layer throughput. The devices are configured with IEEE 802.11ac, MCS 7, short Guard Interval (GI) and 16 dBm of transmission power. The HCM utilizes 80 MHz of BW and 2 SS, while the LCM limits BW to 20 MHz and uses only 1 SS. The current consumption values for the AP's Tx, Rx and Idle states were extracted from real device measurements~\cite{AP_power_consumption}: 1.08 A, 0.66 A, and 0.58 A for LCM; and 1.25 A, 1.00 A, and 0.66 A for HCM, respectively. A Sleep state current of 1.37 $\mu$A is considered, with most AP functions turned off in this state. The considered voltage is 12 V. The results were obtained using the NS-3 network simulator, in a scenario comprising one AP sending DL traffic during 15 s to one non-AP STA.

Since the same throughput can be achieved by different combinations of inter-packet interval and packet size, multiple points, each representing a unique combination of both factors, are shown in Fig.~\ref{fig:pc_throughput} per throughput value. In particular, we varied the inter-packet interval in the range of \(9 \times 10^{-5} \) to \(6 \times 10^{-4} \) seconds and the packet size in the range of 100 to 1500 bytes. The lines represent the linear regression of the points. Results show that LCM is more beneficial for lower throughput values, since idle time is predominant and LCM has the lowest power consumption in this state. However, as throughput increases, idle time is reduced, and it is beneficial to switch to HCM, which provides a higher physical layer data rate and hence lowers the transmission duration, resulting in a reduction in transmit power consumption, the most power-hungry state.
 
Our results show that for this specific scenario and settings, HCM becomes more energy efficient than LCM at a throughput of about 31 Mb/s. The achievable power saving potential is maximal at the lower throughput values, reaching up to 30 percent. Therefore, DPS can be particularly useful at lower loads. BW, transmission power, and NSS have a more significant effect on AP power consumption than MCS. Changing the BW from 20 to 40 MHz, and switching from NSS=1 to NSS=2, increases AP power consumption by 15 percent and by 20 percent, respectively, but changing MCS has a maximum impact of 7 percent \cite{AP_power_consumption}.

Then, in Fig.~\ref{fig:delay_cm}, we show the per-packet delay versus throughput at application layer. In general, there is a trade-off between energy efficiency and QoS. However, as shown in the figure, delay remains mostly unchanged until the network starts to saturate, where it is already more energy-efficient to switch the capability mode to HCM (Fig.~\ref{fig:pc_throughput}). Note that the saturation point for LCM is below 60 Mb/s, so no further values are shown. Also, there is an inherently lower delay for HCM due to its higher data rate. However, if the LCM delay falls within the application requirements, the best option is to remain in the latter mode until the QoS or energy-efficiency requirements are not met.

Additionally, to illustrate the gains of applying SDPS, we conducted an analytical study using real traffic measurements from a public dataset~\cite{ap_dataset}. Said dataset was collected from a total of 470 APs at the University of Oulu, Finland. The measurement period was between December 18, 2018, and February 12, 2019, and each sample provides a 10-minute observation. Fig.~\ref{fig:es_dataset} shows the total traffic, the average power consumption, and the percentage reduction in energy consumption achieved by SDPS over traditional static configurations for 470 APs. We calculate the total traffic (in Mb/s) as the sum of the transmitted (DL) and received (UL) data for every AP and sample, divided by the inter-sample interval. Then, based on the 31 Mb/s threshold from prior experiments, we select the most energy-efficient capability mode. Next, we calculate AP's time in each state (Tx, Rx, Idle, and Sleep), depending on the amount of transmitted and received data, and the chosen mode's data rate. We then determine the average power consumption of each AP by multiplying the time spent in each state by its respective power consumption, and dividing it by the inter-sample interval. For SDPS, if no traffic is exchanged, the AP goes to sleep with a 50 percent probability or remains idle. This decision is derived from~\cite{dynamic_ap_switching}, where authors state that around 50 percent of the APs in a similar scenario can be turned off in off-peak hours with a very low outage probability.

In this scenario, total traffic is noticeably higher during office hours (approximately between 8~a.m. and 8~p.m.), while it is marginal at night, when the highest energy savings occur. Therefore, using SDPS at night can save up to 35 percent compared to a Static configuration, while during office hours, savings are around 15-20 percent, which is still a substantial amount. Over a 24-hour period, SDPS reduces power consumption by 28 percent. Note that these results were obtained using conservative assumptions and a straightforward AP PS approach, as a detailed and complex implementation is beyond the scope of this work. Hence, we anticipate greater savings in future Wi-Fi 8 deployments.

Finally, to demonstrate that AP PS mechanisms can provide substantial power savings across a variety of scenarios, we present results in Table \ref{tab:additional_scenarios} for real Wi-Fi networks in an airport, a cafeteria, and a library, using data from a different public dataset \cite{anonymization_dataset}, along with the results from the studied campus environment. In the new cases, one AP serves a varying number of STAs, with traffic captured over twenty-minute periods at different times of the day. As with Fig. \ref{fig:es_dataset}, we calculate the percentage of power savings with DPS and SDPS to highlight the impact of enabling energy-saving mechanisms. With SDPS, the AP transitions to sleep state during periods of inactivity with a 50 percent probability (~\cite{dynamic_ap_switching}). The results show power savings of around 10-12 percent with DPS, increasing up to 13-28 percent with SDPS, while maintaining network service. Table \ref{tab:additional_scenarios} also presents the APs' average percentage of busy time in 1-second intervals, which indicate prevalent low-throughput and idle periods, suggesting significant power-saving potential, even during busy hours. Note that these results represent a lower bound, as the dataset \cite{anonymization_dataset} does not include low-traffic periods (e.g., nighttime), which could provide further power-saving opportunities. For more detailed information on the configuration of these scenarios, please refer to \cite{anonymization_dataset}.

\begin{table}[t]
\caption{Power saving for DPS and SDPS, for a variety of scenarios}
\centering
\begin{tabularx}{\linewidth}{|>{\centering\arraybackslash\color{black}}X|>{\centering\arraybackslash\color{black}}X|>{\centering\arraybackslash\color{black}}X|>{\centering\arraybackslash\color{black}}X|}
\hline
\textbf{Scenario}&\textbf{Busy time (\%)}&\textbf{Power saving with DPS (\%)}&\textbf{Power saving with SDPS (\%)}\\
\hline
Airport & 0.761 & 9.90 & 17.51 \\
\hline
Cafeteria & 0.128 & 11.98 & 12.96 \\
\hline
Library & 0.045 & 12.25 & 20.41 \\
\hline
Campus & 0.007 & 12.05 & 27.91 \\
\hline
\end{tabularx}
\label{tab:additional_scenarios}
\end{table}

\section{Discussion and Open challenges}

In this article, we outlined AP PS mechanisms that are under TGbn standardization, each with advantages and limitations (Table~\ref{tab:characteristics}). Scheduled PS is ideal for applications with a periodic nature, but it is not adaptive due to its pre-scheduled nature. In contrast, (S)DPS provides a dynamic change of capabilities, allowing for a quick adaptability to different traffic profiles, but it can introduce unfairness with legacy devices. Performance can increase when combined with Scheduled PS, although it can introduce unfairness and inability for the AP to switch to doze state. Cross-Link excels for traffic profiles with sudden bursts of time-critical data, although it suffers from a significant lack of backward compatibility, as does WuR, which suits for very infrequent data exchanges. Finally, STA offloading is a useful solution to save energy for an extended period of time, however, it can introduce significant delay when traffic characteristics change rapidly. Ultimately, the choice among these mechanisms depends on factors such as desired power saving, adaptability, and backward compatibility, and requires a trade-off between energy efficiency and QoS. A hybrid, adaptive approach combining all or a subset of the mechanisms, depending on the scenario, would offer the best results in both energy efficiency and QoS.

The sections below overview the key challenges that need to be addressed for AP PS, namely resource allocation, backward compatibility, and signaling overhead.

\setlength{\tymin}{2cm}

\begin{table*}[t]
\caption{Summary, signaling, power saving, schedule duration, backward compatibility, and recommended application scenarios} for the proposed AP PS mechanisms.
\centering
\begin{tabulary}{\textwidth}{|L|L|L|L|L|L|L|}
\hline
\textbf{AP PS mechanism}&\textbf{Summary}&\textbf{Signaling}&\textbf{Power saving}&\textbf{Schedule duration}&\textbf{Backward compatibility}&\textbf{Recommended application scenarios}\\
\hline
\textbf{Scheduled} & Scheduling of SPs with different operational modes & Reuse TWT signaling, presence requests & High & Long-term & Low since legacy devices require AP to be in awake state & Traffic profiles with a periodic nature (e.g., periodic IoT sensors)\\
\hline
\textbf{Dynamic (DPS)} & On-demand switching of capabilities & ICF with intermediate FCS and padding, ICR
& Medium & Short-term & Limited due to unfairness, because legacy devices use mostly LCM & Unpredictable traffic profiles (e.g., spontaneous video conferencing, streaming) \\
\hline
\textbf{Semi-Dynamic (SDPS)} & DPS modification, when AP can defer switching. Can be combined with scheduled. & Same as for Scheduled and DPS, LL flag added to ICF & Medium to high & Both short and long-term & Limited due to unfairness and inability for the AP to switch into doze state & Dynamic traffic profiles where both high data rate and low latency are required (e.g., low-latency streaming, interactive features, online gaming) \\
\hline
\textbf{Cross-Link} & On-demand wake-up other links via an active link of MLD AP & Cross-Link wake-up (reuse AP assistance request) & High & Short-term & High since communication via active link is available & Sudden bursts of data (e.g., reports from/to sensors/actuators in industrial environments, alarms in emergency situations) \\
\hline
\textbf{Wake-up Radios (WuRs)} & On-demand wake-up of primary radio via companion WuR & IEEE 802.11ba signaling & Very high & Short-term & Low since WuRs are not supported by legacy devices & Very infrequent data exchanges (e.g., sporadic or event-based sensor data collection) \\
\hline
\textbf{STA offloading} & Offloading of STAs to other APs to save energy & Seamless roaming signaling via Multi-AP Coordination & High & Long-term & High through legacy handover procedures (although with worse performance) & Scenarios with traffic seasonality over a long interval, such as during a day (e.g., exhibition or institutional buildings) \\
\hline
\end{tabulary}
\label{tab:characteristics}
\end{table*}

\subsection{Resource allocation}

The PS mechanisms add several degrees of freedom regarding the allocation of channel resources. Specifically, the AP should schedule SPs and PS periods and assign traffic flows to specific capability modes and links. According to SDPS, STAs trigger the AP to switch the capabilities, but the mapping between the QoS requirements of different traffic flows and the capability modes at the STA side seems non-trivial. Moreover, for each request, the AP can decide to accept or decline it depending on many factors, such as load, requirements of ongoing flows, channel quality per STA, and others. The needed resource allocation algorithms are expected to be out of scope of the standard, and thus, contributions from the research community will be necessary.

\subsection{Compatibility with legacy devices}

In controlled environments, such as industrial deployments, the devices used for connectivity are known beforehand. Maintainers of such deployments can ensure that all Wi-Fi devices support the same features. However, in many scenarios this is not the case. In such scenarios, it is important that new features do not cause significant performance degradation for legacy devices. With respect to AP PS, the following issues should be taken into account. First, the AP cannot inform legacy devices about going into doze state. Although the AP could rely on legacy Quiet Elements to forbid the STAs' transmissions during the doze period, experimental studies revealed inconsistency in the implementation of this mechanism in legacy devices~\cite{quieting}. Besides, since this is an optional feature, the AP will restrict access to STAs supporting IEEE 802.11h. Second, legacy devices are not able to handle the PS schedule from the AP and cannot trigger the AP to switch capabilities. These issues cause several potential complications, such as: 

\begin{itemize}
    \item The AP never goes into doze state.
    \item Performance unfairness between IEEE~802.11bn and legacy devices, as legacy devices are primarily limited to LCM for transmission.
    \item The AP offloads all legacy devices to other APs to maximize the benefits from PS features, which may lead to those other, probably legacy, APs becoming overloaded.
\end{itemize}

\subsection{Signaling overhead}

Every new feature introduced in the standard requires additional signaling. Often, for backward compatibility and to limit overhead, the frames used for already existing procedures and mechanisms are reused for the new ones. For example, many control frames used in previous Wi-Fi generations have sub-fields that were intentionally reserved for future use. Current discussions in TGbn indicate plans to reuse TWT signaling for Scheduled PS; Buffer Status Report Poll (BSRP) or Multi-User Ready To Send (MU-RTS) for DPS; and AAR for Cross-Link PS. However, since these frames have been initially designed for other procedures, their application to AP PS will require changes both at STA and AP sides. Besides, the current Wi-Fi standard does not have a common framework for exchanging capability information for PS features. Proposals for a modular and extensible framework for capabilities exchange are currently being discussed in TGbn. In any case, new signaling would introduce overhead and, consequently, delays related to transmission of new control frames, switching between capabilities, enabling/disabling links. The effect of this overhead should be studied.

\section{Conclusion}

Energy consumption is an increasing concern in the ICT sector in general and in Wi-Fi networks in particular, both due to the associated costs and carbon emissions. In this article, we outlined the new AP PS framework, under consideration for IEEE 802.11bn standardization. Specifically, we considered six different mechanisms that are being discussed in TGbn: Scheduled, Dynamic, Semi-Dynamic, Cross-Link, WuR and STA offloading. For each of the proposals we highlighted their operation flow, used signaling, potential benefits and issues. Finally, we provided a case study to quantitatively assess the potential energy savings that can be achieved in a real university campus when making use of AP PS. The results are based on conservative assumptions from a proof-of-concept AP PS approach, with greater savings expected in future Wi-Fi 8 deployments.

\section*{Acknowledgments}
This research was funded in part by the Spanish MCIU/AEI/10.13039/501100011033/FEDER/UE through projects PID2019-106808RA-I00 and PID2023-146378NB-I00, and by Secretaria d’Universitats i Recerca del departament d’Empresa i Coneixement de la Generalitat de Catalunya with the grant number 2021 SGR 00330. The~first author gratefully acknowledges the predoctoral program AGAUR-FI grant (2023 FI-1 00154) Joan Oró of the Secretaria d'Universitats i Recerca del Departament de Recerca i Universitats de la Generalitat de Catalunya and the European Social Fund Plus.

%\bibliographystyle{IEEEtran}
%\bibliography{bibliography}

\section{Biographies}

\vspace{11pt}

\begin{IEEEbiographynophoto}{Roger Sanchez-Vital}
Roger Sanchez-Vital received his M.Sc. in Telecommunications Engineering from La Salle–Universitat Ramon Llull in 2021 and is now pursuing a Ph.D. in Network Engineering from the Universitat Politècnica de Catalunya (UPC). He is a predoctoral researcher at the same institution. His research interests mainly include: IoT, LPWAN technologies, and the IEEE 802.11 standard.
\end{IEEEbiographynophoto}

\begin{IEEEbiographynophoto}{Andrey Belogaev}
Andrey Belogaev is a senior researcher at the University of Antwerp and imec, Belgium. His research interests lie in the protocol design and optimization for future Wi-Fi and cellular networks. He obtained his M.Sc. (2018) and Ph.D. (2020) from Moscow Institute of Physics and Technology.
\end{IEEEbiographynophoto}

\begin{IEEEbiographynophoto}{Carles Gomez}
Dr. Carles Gomez obtained his Ph.D. thesis in Network Engineering in 2007. He is a Full Professor at the Universitat Politècnica de Catalunya (UPC). He is a co-author of numerous papers published in journals and conferences, as well as IETF RFCs and Internet Drafts. He is an IETF 6Lo Working Group Chair. He serves as an editorial board member for several journals. His research interests mainly focus on the Internet of Things and deep space communication.
\end{IEEEbiographynophoto}

\begin{IEEEbiographynophoto}{Jeroen Famaey}
Jeroen Famaey is a research professor at the IDLab research group of the University of Antwerp and imec, Belgium. At IDLab, he leads the Connected Systems research team, which focuses on signal processing, protocol design, and management of wireless communication systems. His research has led to the publication of over 180 peer-reviewed articles in journals and conference proceedings, and 8 granted patents.
\end{IEEEbiographynophoto}

\begin{IEEEbiographynophoto}{Eduard Garcia-Villegas}
Dr. E. Garcia-Villegas is an Associate Professor at the Universitat Politècnica de Catalunya (UPC), where he specializes in radio resource optimization, network security, the IoT, and emerging 5G/6G technologies. Holding a Ph.D. in Telecommunications Engineering, Dr. Garcia-Villegas has made significant contributions to Wi-Fi technology and has actively participated in several IEEE P802.11 task groups. As a member of the Wireless Networks Group (WNG) and the i2CAT Foundation, he collaborates extensively with both industry and academia on pioneering wireless network initiatives.
\end{IEEEbiographynophoto}

\vfill

\end{document}